\documentclass[twocolumn,amssymb,fleqn, twocolumns]{revtex4} 
\usepackage{epsfig,amssymb,amsmath,graphicx,subfigure,hyperref  }  
\usepackage{color}

\usepackage{multirow} 
\usepackage{tabularx}

\begin{document}

\title{Charged elastic rings: deformation and dynamics}
\author{Zhenwei Yao}
\email[]{zyao@sjtu.edu.cn}
\affiliation{School of Physics and Astronomy, and Institute of Natural Sciences, Shanghai Jiao Tong University, Shanghai 200240, China}
\begin{abstract} 
  We report the counter-intuitive instability of charged elastic rings, and the
  persistence of sinusoidal deformations in the lowest-energy configurations by
  the combination of high-precision numerical simulations and analytical
  perturbation calculation. We also study the dynamical evolution of the
  charged ring under random disturbance, and reveal the modulation of
  the dominant frequencies by the electrostatic force. The purely mechanical
  analysis of the classical ring system presented in this work yields insights
  into the subtlety of long-range forces in the organization and dynamics of
  matter. 
\end{abstract}

\maketitle

\section{Introduction}

As a fundamental force of nature, the electrostatic force is highly involved in
a myriad of important physical
processes~\cite{roller1970development,schrack1985electrical,levin2005strange,vernizzi2007faceting,panagiotopoulos2009charge}.  Complex and even
counter-intuitive phenomena arise when electrostatics is combined with
many-body physics~\cite{messina2008electrostatics,Walker2011}.  Examples include the reversal of
charges~\cite{torrie1980electrical,pianegonda2005charge,Grosberg2002} and the
attraction of
like-charges~\cite{Bloomfield1991,Gelbart2000,Levin2002,Walker2011,qin2016theory} in electrolyte solutions, and the unexpected packings of classical
point charges on a disk~\cite{berezin1985unexpected,macgowan1985electrostatic}. Especially, by low-dimensional classical model
systems, much has been learned about the organization of point charges of the
same sign that dislike each
other~\cite{Bausch2003e,irvine2010pleats,soni2018emergent,yao2019command}. Notably, the study of the 100-year-old
  Thomson problem~\cite{thomson1904xxiv}, which aims at finding the ground state
configuration of point charges confined on the sphere, yields rich insights into
a host of fundamental questions, including the physics of topological defects
and crystallography on curved
space~\cite{bowick2002crystalline,bowick2009two,wales2014chemistry}, the self-assembly of
virus~\cite{chiu1997structural,lidmar2003virus}, and the development of
relevant algorithms~\cite{altschuler1997possible,mehta2016kinetic}. In spite of
its simplicity, the Thomson problem has not been solved yet.

In this work, we explore an even simpler one-dimensional system, where a
collection of point charges are connected by linear springs. The one-dimensional
problem of the equilibrium distribution of point charges on frozen geometries
has been first discussed by Maxwell~\cite{maxwell1877electrical}, and later
rediscovered by several
authors~\cite{griffiths1996charge,jackson2000charge,jackson2002charge,amore2019thomson}.
By including elasticity, which represents the simplest organization of matter,
the electrically charged elastic ring system possesses the essential elements
of electrostatics and many-body physics, and it serves as a suitable model to
address the inquiry into the optimal organization and collective dynamics of
long-range repulsive particles. The ring model also has connections with a
variety of loop polymers and knotted
biopolymers~\cite{dommersnes2002knots,fortini2005phase,cherstvy2011dna,lin2015pulled,huang2019shape}.

The main results of this work are presented below. We first resort to
high-precision numerical simulations to explore the lowest-energy configurations
of charged elastic rings under the competing electrostatic and elastic forces in
three-dimensional space, and reveal the counter-intuitive instability of the
ring system. The persistence of sinusoidal deformations is identified in the
lowest-energy configurations. The long-range nature of the interaction potential
is crucial for the subtle instability of the ring system. We further
substantiate the numerical results by analytical perturbation calculation in
the continuum limit. The combination of the numerical results and the analytical
perturbation analysis shows that the ground state shape of the charged elastic
ring is not a perfect circle. While the non-circular shape has been 
identified as the lowest-energy state by steepest descent method under high
precision, it is still an open question to rigorously prove if this elliptic shape is the
ground state of the system.

We further investigate the dynamical effect of the electrostatic force by
examining the dynamical evolution of the ring upon random disturbance. By
spectral analysis of the temporally-varying kinetic energy curves, we reveal a
pair of dominant frequencies, and identify the corresponding dynamical modes. We
also show the modulation of the dominant frequencies by increasing the
strength of the electrostatic force. This work advances our understanding on the
crucial role of long-range electrostatic force in the equilibrium organization
and collective dynamics of many-body systems.

\section{Model and Method}

\begin{figure}[t]  
\centering 
\includegraphics[width=3.3in]{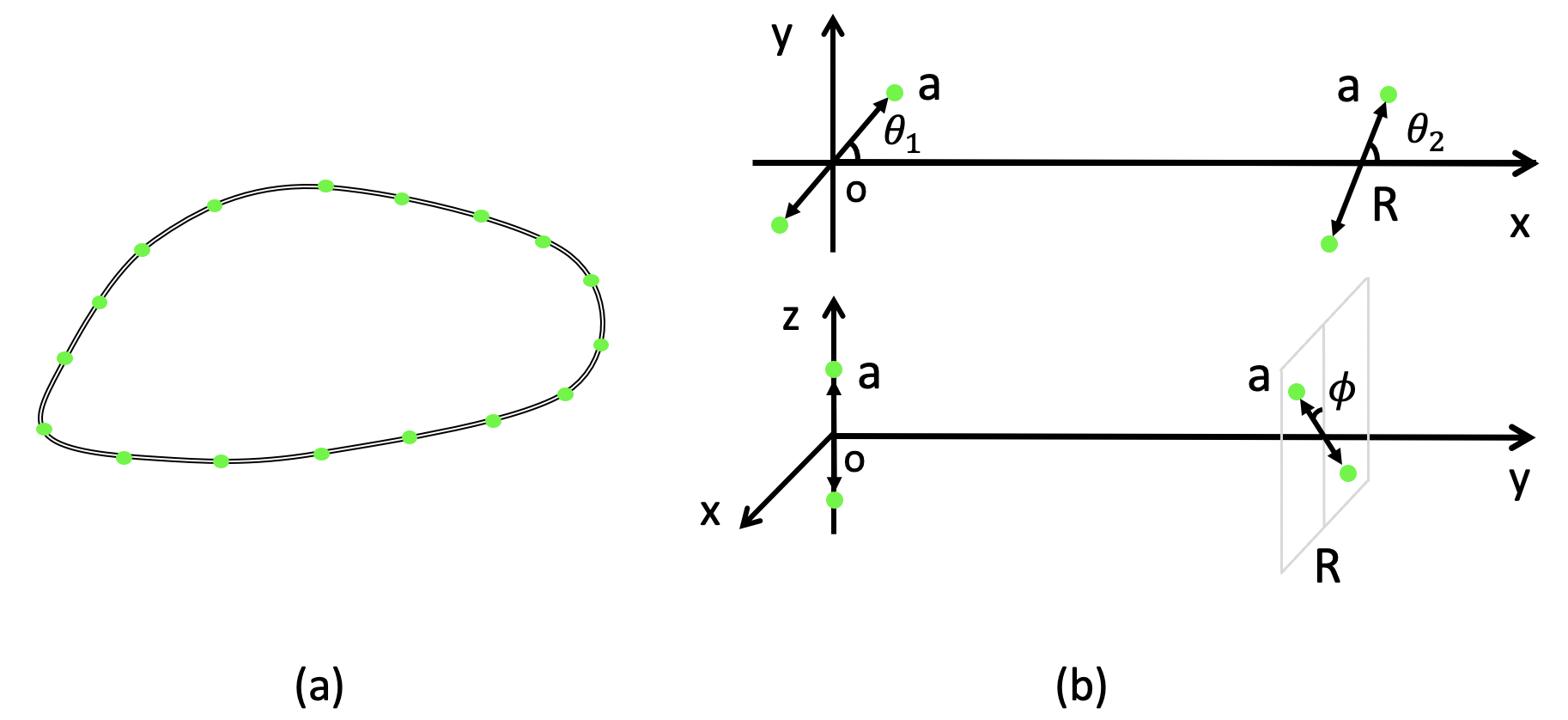}
  \caption{Schematic plot of the model system. (a) The model consists of a
  collection of point charges (green dots) connected by linear springs (black
  lines). (b) The model systems of charged pairs for energetic analysis (see
  more information in the text).  }
\label{schematic}
\end{figure}

Our model consists of $N$ point charges connected by linear springs as
shown in figure~\ref{schematic}(a). The potential energy of the system
is 
\begin{align}
  E_p = \sum_{j>i} \frac{\beta}{r_{ij}} + \sum_{\alpha} \frac{1}{2}k_0
  (\ell_{\alpha}-\ell_0)^2, \label{Ep}
\end{align}
where the first and second terms are the electrostatic and elastic energy,
respectively. $\beta=q_0^2 /(4\pi \varepsilon)$, where $q_0$ is the charge on
each particle, and $\varepsilon$ is the dielectric constant of the medium where
the ring is immersed. $r_{ij}$ is the Euclidean distance between the particles
$i$ and $j$. $k_0$ is the spring stiffness, $\ell_{\alpha}$ is the actual length
of the spring $\alpha$ and $\ell_0$ is the balance length. In connection with
real systems, our model contains the common elements of electrostatics and
elasticity in a variety of soft matter systems in electrolyte
environment~\cite{Walker2011}. Especially, many biopolymers are highly charged,
and the formation of tight knots under the interplay of Coulomb interaction and
topological constraints has been reported~\cite{dommersnes2002knots}. The
Coulomb potential used in our model represents the limiting case of large Debye
length in dilute solutions.  In this work, the units of length, energy and force
are $\ell_0$, $\epsilon_0$, and $f_0$, where $\epsilon_0 = k_0 \ell_0^2$, and
$f_0=\epsilon_0/\ell_0$.

We employ high-precision numerical simulations to search for the lowest-energy
shape of the electrically charged elastic ring system. The initial configuration
of the ring is a circle with imposed random undulation in 3D space. By the
steepest descent method, the particle configuration is updated under decreasing
step size $s$ over several million simulation cycles. Each simulation cycle
consists of a collective update of the particle positions by the force vector on
each particle. 

To further study the dynamical effect of the electrostatic force, we impose
random disturbance to the ring configuration in mechanical equilibrium. The
subsequent motion of the ring conforms to the classical Hamiltonian dynamics.
Specifically, as the initial state, each particle in the lowest-energy
configuration of the ring is specified by a random velocity $\vec{v}_{ini}$.
The $\alpha$-component $v_{ini, \alpha} \in [-v_0, v_0]$, where $\alpha=x, y,
z$.  The strength of the disturbance is characterized by the value of $v_0$. The
subsequent dynamical evolution is governed by the Hamiltonian 
\begin{align}
  H=\sum_{i=1}^{N}\frac{p_i^2}{2m} + E_p,
\end{align} 
where the first term is the total kinetic energy, and the total potential energy
$E_p$ is given in equation~(\ref{Ep}).  We adopt Verlet method to construct
energy-conserved particle trajectories~\cite{rapaport2004art}. In numerical
simulations, the time step $\delta t$ is chosen to be sufficiently fine to
ensure the conservation of the total energy. Typically, $\delta t$ is at the
order of $10^{-4} \tau_0$ and $10^{-5} \tau_0$, where $\tau_0=\ell_0
\sqrt{m/\epsilon_0}$.

\section{Results and Discussion}

\subsection{Preliminary analysis}

The mechanical relaxation of the electrically charged elastic ring system is
dominated by the competing electrostatic and elastic forces. Under the
long-range electrostatic force, the point charges tend to be uniformly
distributed over an expanding sphere in three-dimensional space, which is
suppressed by the elastic force. To investigate the ground state shape of the
charged elastic ring, we first analyze the interaction between elementary charge
pairs along the ring.

A charge pair consists of two point charges of the same sign separated by $a$.
For two charge pairs at distance $R$ in the x-y plane, as shown in the upper
panel in figure~\ref{schematic}(b), the total electrostatic energy is
\begin{align}
  U(\theta_1, \theta_2) = \frac{2\beta}{R}\left[ 2+\tilde{a}^2(4-3 (\sin^2\theta_1 + \sin^2
  \theta_2)) \right], \label{U1}
\end{align}
where $\tilde{a}=a/R$. From equation~(\ref{U1}), we see that the two charge pairs tend
to be parallel to each other and perpendicular to the connecting line for minimizing the
interaction energy. For the three-dimensional case, where one charge pair is
fixed and the other charge pair can freely rotate in the x-z plane [see the
lower panel in figure~\ref{schematic}(b)], the total electrostatic energy is
\begin{align}
  U(\phi)=\frac{2\beta}{R} \left[ 2(1-\tilde{a}^2) + 3 \tilde{a}^4 (1+\cos^2\phi)
  \right]. \label{U2}
\end{align}
Equation (\ref{U2}) shows that the charge pair at $y=R$ tends to be perpendicular to
the other charge pair by taking $\phi=\pi/2$ for minimizing the interaction energy. The
analysis of equations~(\ref{U1}) and (\ref{U2}) suggests that a circular charged ring
might be unstable. 

Now, we discuss the electrostatic force on a closed
charged ring in the continuum limit. Variational calculation shows
that the electrostatic force is always along the principal normal
direction~\cite{goldstein1995nonlinear,santiago2013elastic}. Specifically, the
energy functional associated with the electrostatics of a closed, uniformly
charged space curve $\vec{r}(s)$ is
\begin{align}
  \mathcal{E}=\frac{1}{2}\oint ds \oint ds' \Phi(|\vec{r}(s)-\vec{r}(s')|).
  \end{align} 
The corresponding electrostatic force 
\begin{align}
  F_{e} = -(\kappa(s) + \hat{n}\cdot \nabla)\oint ds' \Phi\left(r(s, s')\right), \label{Fe}
\end{align}
where $\kappa(s)$ is the curvature, $\hat{n}$ is the principal normal vector,
and $r(s,s')=|\vec{r}(s)-\vec{r}(s')|$~\cite{goldstein1995nonlinear,santiago2013elastic}.
The electrostatic force acting on the length element at $\vec{r}(s)$ is along
the normal direction $\hat{n}(s)$.  Note that equation~(\ref{Fe}) is valid for any
pair interaction potential $\Phi(r)$.  While the purely normal electrostatic
force could support the planar circular shape with constant curvature, it is
unknown if the circular ring is stable, especially by the interplay of
electrostatics and elasticity.

\subsection{Deformation of the ring}

To explore the lowest-energy configuration of charged elastic rings under the
competing electrostatic and elastic forces in three-dimensional space, we resort
to numerical simulations at high precision based on the steepest descent method.
Numerical simulation has proven to be a suitable tool to investigate the
mechanically equilibrium properties and dynamics of long-range interacting
particle systems~\cite{yao2016electrostatics,yao2021epl}.

\begin{figure}[t]  
  \centering
  \includegraphics[width=3.4in]{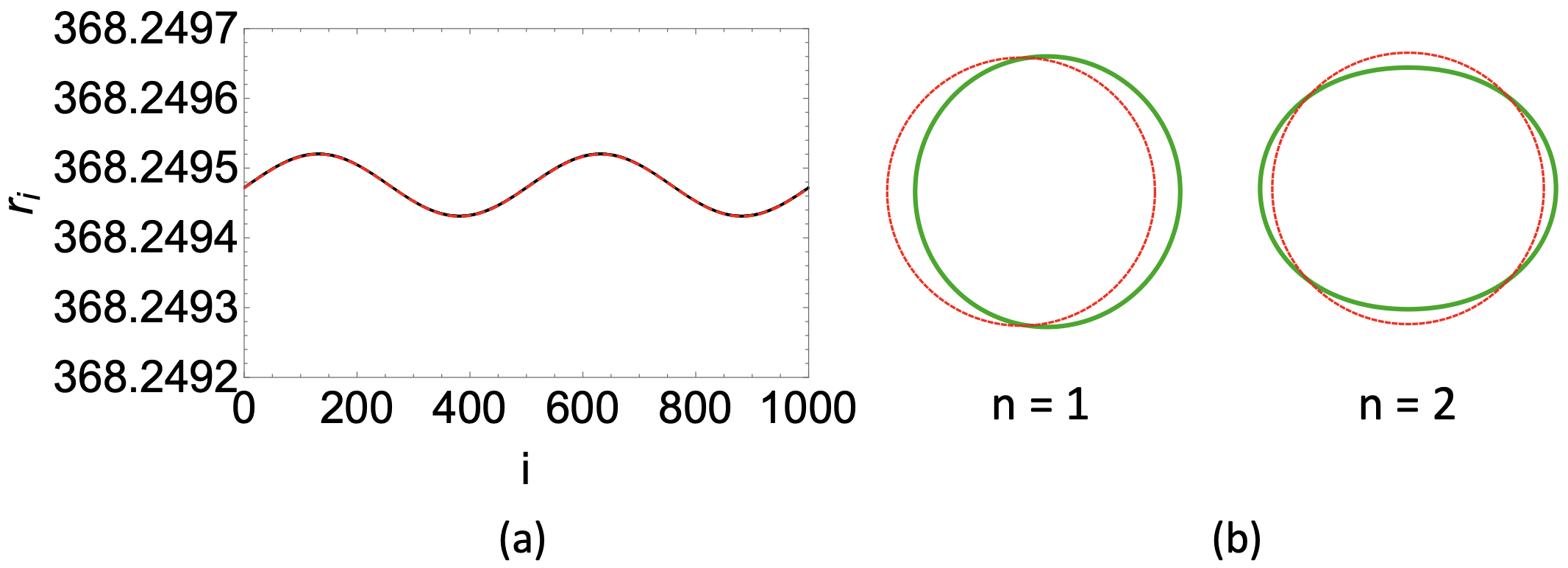} 
  \caption{The numerically obtained lowest-energy shape of the electrically
  charged, elastic ring system by steepest descent method under high precision.
  (a) Plot of the distance $r_i$ from each particle to the center of mass of the
  ring. The $r_i$ curve could be precisely fitted by the shape of the $n=2$
  mode, which is indicated by the dashed red fitting curve. The step size $s$ is
  reduced to as low as $10^{-8}$ in simulations.  (b) The shapes of the $n=1$
  and $n=2$ modes. $n$ indicates the deformation mode; see equation~(\ref{n_mode}).
  The reference circles (dashed, red) are also shown.  $N=1000$.  $\beta=1$.
  }
\label{deformation_final} 
\end{figure}

Preliminary numerical simulations show that the deformed rings in
three-dimensional space are ultimately flattened to circles under the combined
electrostatic and elastic forces. However, closer examination shows the
existence of slight, but persistent sinusoidal deformations on the rings.
Notably, in the long-time relaxation process, wave structure is developed on the
originally circular ring. The wave number keeps reducing as the system evolves
towards the lowest-energy shape. Simulation details are provided in SI.  Note
that the wave structure on the ring may be stabilized by introducing the
attribute of spontaneous curvature to the ring
system~\cite{hossein2020spontaneous}. The undulated shapes also provide clues
for the ring geometry due to both thermal fluctuations and combination of the
electrostatic and elastic interactions.  In figure~\ref{deformation_final}(a),
we plot the distance $r_i$ from each particle to the center of mass for the
numerically obtained lowest-energy shape under the precision of step size
$s=10^{-8}$. The $r_i$ curve could be precisely fitted by the shape of the $n=2$
mode: $r(\theta) = r_1 + h \cos(2\theta+\theta_0)$. Both shapes of the $n=1$ and
$n=2$ modes are shown in figure~\ref{deformation_final}(b); the reference
circles (dashed, red) are also shown. The relative deviation of the fitting
function and the data, which is measured by $\sum_{i=1}^N
|r_i-r(\theta_i)|/\langle r_i \rangle$, is as small as $10^{-10}$.

The shape of the $n=1$ mode is absent in numerical simulations, because this
mode essentially describes the translation of the entire ring in the
perturbation regime. For the $n=1$ mode, the translation $\delta x$ along the
x-axis satisfies the relation $(r_1+h_0\cos\theta-r_0)/\delta x = \cos\theta$,
which leads to $\delta x = h_0+O(h_0^2)$.  It is only for the $n=1$ mode that
the linear term of $h_0$ in the expression for $\delta x$ is independent of
$\theta$. As such, the $n=2$ mode represents the first deformation mode in the
perturbation regime that breaks the continuous rotational symmetry of the
circular ring.

We further quantify the degree of the out-of-plane deformation by the mean squared
distance $\Delta$ of the particles with respect to the reference plane as spanned
by the three particles $i_1=1$, $i_2={\textrm
{floor}}(N/3)$, and $i_3={\textrm {floor}}(2N/3)$.  
\begin{align}
  \Delta  = \frac{1}{\ell_{ext}}\sqrt{\frac{\sum_{i=1}^{N} d_i^2}{N}}.
\end{align}
where $d_i$ is the distance from the particle $i$ to the reference plane.
$\ell_{ext}$ is the lateral extension of the deformed ring, which is equal to
the mean side length of the reference triangle. Simulations show that the
magnitude of the out-of-plane deformation is much smaller than that of the
sinusoidal deformation by two orders of magnitude; the detailed information is
provided in SI. In other words, the sinusoidal deformation occurs approximately
in the plane. It is further observed that the standard deviation of the bond
length distribution is much smaller than the amplitude of the sinusoidal
deformation by two orders of magnitude.

\begin{figure}[t]  
\centering 
\includegraphics[width=3in]{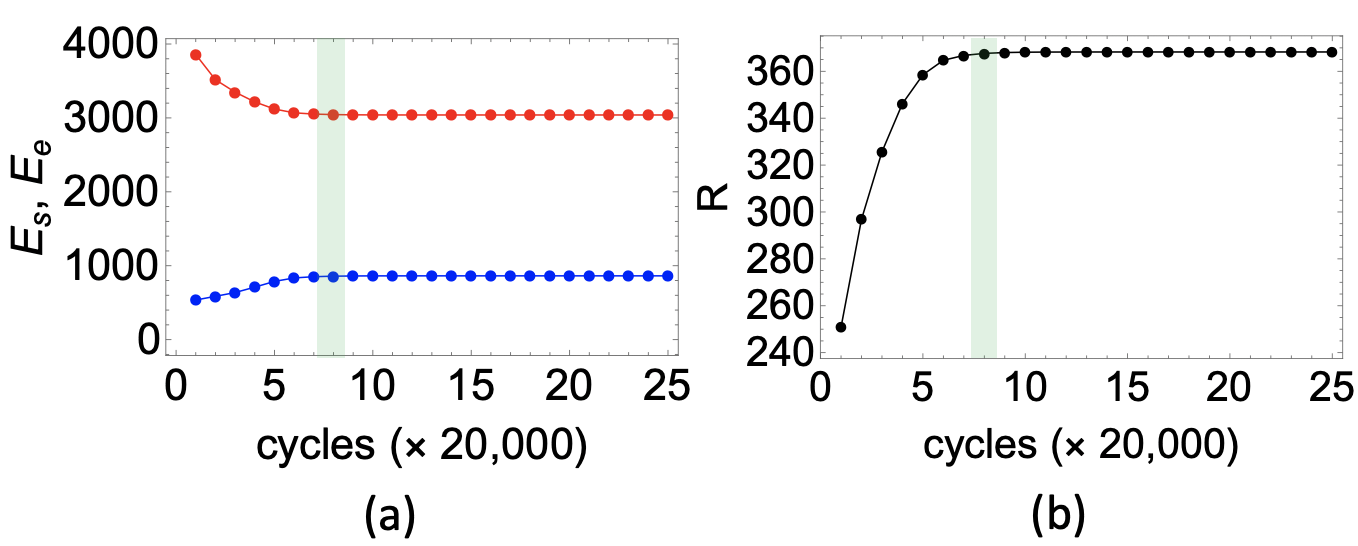}
  \caption{Variation of energy and ring size in the simulated relaxation
  process. (a) $E_s$ and $E_e$ are the stretching and electrostatic energies. (b) $R$ is
  the mean radius of the ring. $N=1000$. $\beta=1$.  }
\label{energetics}
\end{figure}

Systematic numerical simulations show that the persistence of the quasi-planar
sinusoidal deformation in the lowest-energy configurations is a common
phenomenon for varying $N$ and $\beta$. We examine typical cases in the parameter
space of $N\in [100, 4000]$ and $\beta \in [10^{-5}, 10^6]$ by high-precision
simulations, and observe that the lowest-energy configurations of the charged
elastic rings uniformly converge to the shape of the $n=2$ mode; the maximum
residual force on the particles is typically at the order of $10^{-7}f_0$. We
shall emphasize that the amplitude of the sinusoidal deformation is much smaller
than the mean radius of the ring. For example, analysis of the lowest-energy
configurations as obtained for the step size $s$ at the order of $10^{-4}$ shows
that the ratio of the amplitude of the sinusoidal deformation and the mean
radius is at the order of $10^{-5}$ when the value of $\beta$ is changed by
three orders of magnitude from unity to 1000. As such, it is a challenge to
experimentally observe the sinusoidal deformation in the lowest-energy charged
elastic ring. The exclusion of thermal fluctuations and acoustic waves in
numerical simulations allows us to reveal the sinusoidal deformation.

In the electrostatics-driven deformation of the charged elastic ring, the
electrostatic and elastic energies have opposite dependence on the system size.
As such, the ring size ultimately reaches an equilibrium value in the relaxation
process. The variations of energy and ring size in the relaxation process are
plotted in figure~\ref{energetics}(a) and~\ref{energetics}(b), respectively.  In
the following, we perform scaling analysis of the equilibrium size of the ring.
The asymptotic expression for the electrostatic energy of $N$ point charges
evenly distributed on a circle of radius $R$ in the large $N$ limit
is~\cite{Saff_asymptoticsfor,Borodachov2012}
\begin{align}
  E_{{\textrm {e}}} \propto \frac{1}{R} N^2\ln N. \label{Escale}
\end{align}
The elastic stretching energy 
\begin{align} 
  E_{{\textrm {s}}} \propto \frac{1}{N} (R-R_0)^2,
  \end{align}
where $R_0=N\ell_0$. $\ell_0$ is the balance length of the spring. From $\delta
E_{{\textrm {tot}}}/\delta R|_{{\textrm {equil.}}  }=0$, we obtain the scaling
law:
\begin{align} 
  R_{{\textrm {equil.}}} \propto N(\ln N)^{1/3}. \label{Req}
\end{align}
Equation~(\ref{Req}) is confirmed in our numerical simulations (see SI).

Here, it is of interest to discuss the division of a charged ring
into two identical rings based on equation~(\ref{Escale}); the perimeter of the ring is
conserved in the division process. Without considering the interaction of the
two rings, the change of the electrostatic energy is
\begin{align} 
  \frac{\Delta E}{E_0} = -\frac{\ln 2}{\ln N} < 0,
\end{align}
which indicates the reduction of the electrostatic energy as the ring is equally
divided. This result is the two-dimensional version of the electrostatics-driven
instability of charged spheres in the seminal work by
Rayleigh~\cite{rayleigh1882xx}.

\subsection{Perturbation analysis in the continuum limit}

In this subsection, we perform analytical perturbation calculation to further
substantiate the main conclusion of the preceding numerical analysis that the
ground state shape of the charged elastic ring is not a perfect circle as one
may expect intuitively. 

According to the numerical results, both the out-of-plane deformation and the
standard deviation of the bond length distribution in the lowest-energy
configurations are vanishingly small. As such, we consider the configurations of
uniformly distributed particles on a perfect circle and on a slightly deformed
ring in the plane. Here, for the sake of analytical tractability as well as
considering the numerical results, the bond length on both configurations is set
to be identical, so their elastic energies are the same. The variation of the
energy due to the deformation of the ring is completely determined by the
electrostatic part. Could the electrostatic energy be lowered by deforming the
uniformly charged circular ring? This question is also closely related to the stability
of a uniformly charged circular ring.

In the following, we perform perturbation analysis in the continuum limit to address
this question. According to the preceding discussion, the linear charge density
is invariant in the deformation, and the deformation is confined in the plane.
Now, consider a slight deformation of the uniformly charged ring in the plane:
\begin{align}
  \vec{r}(\theta) = \vec{r}_0 + h(\theta) \hat{n}(\theta),
\end{align}
where $\hat{n}(\theta)$ is the outward unit normal vector on the original
circular ring of radius $r_0$. $\vec{r}_0 = r_0 \hat{n}(\theta)$. The resulting change of the total electrostatic energy is
\begin{align}
  \Delta E = \lambda^2 \oint  d\ell \oint d\ell' V(r_{pp'}) - 
  \lambda^2 \oint d\ell_0 \oint d\ell'_0 V(r_{p_0 p_0'}),  \label{DE1}
\end{align}
where the first and the second terms are the integrals over the perturbed and
circularly distributed charges. $\lambda$ is the linear charge density. $V(r)$
is the interaction potential of two unit charges separated by distance $r$.
$d\ell$ and $d\ell'$ are the line elements at any two points $p$ and $p'$ on the
perturbed ring. And $d\ell_0$ and $d\ell'_0$ are the line elements at any two
points $p_0$ and $p'_0$ on the circular ring.

In the following, we expand the expression of $\Delta E$
in terms of the perturbation $h(\theta)$. We first expand $|\vec{r}_{pp'}|$,
which appears in the first term in equation~(\ref{DE1}), in terms of $h$. The position vector
connecting the points $p$ and $p'$:
\begin{align}
  \vec{r}_{pp'} = \vec{r}_{p_0p_0'} + \delta \vec{r}, 
\end{align}
where $\delta \vec{r} = h(\theta')\hat{n}(\theta') - h(\theta)\hat{n}(\theta)$.
By using the following relation: 
\begin{align}
  \left| \vec{f}+\delta \vec{f} \,\right| = f+\Delta f, 
\end{align}
where $f=|\vec{f}\,|$ and 
\begin{align}
\Delta f=\frac{\vec{f}\cdot \delta\vec{f}}{f} +\frac{1}{2}\frac{\delta \vec{f}\cdot \delta \vec{f}}{f}
-\frac{1}{2}\frac{(\vec{f}\cdot\delta \vec{f})^2}{f^3}+O(\delta f^3), \nonumber
\\
\end{align} 
we have 
\begin{align}
  r_{pp'} = r_{p_0p_0'} + \Delta r,
\end{align}
where
\begin{align}
  r_{p_0p_0'}= 2r_0 \biggr | \sin\frac{\theta-\theta'}{2} \biggr | ,
\end{align}
and 
\begin{align}
\Delta r= \frac{r_{p_0p_0'}}{2r_0} \left( h(\theta) + h(\theta')
  \right) + O(h^2). \label{Dr}
\end{align}
Now, we insert the following equations to equation~(\ref{DE1}):
\begin{align}
  d\ell =  r_0\left( 1+\frac{h(\theta)}{r_0} + \frac{1}{2}
  \frac{h'(\theta)^2}{r_0^2} + O(h^3) \right) d\theta, \label{dell}
\end{align}
and 
\begin{align}
  V(r_{pp'}) = V(r_{p_0p_0'}) &+& \left. \frac{dV(x)}{dx} \right|_{x=r_{p_0p_0'}} \Delta r \nonumber \\
  &+& \frac{1}{2} \left. \frac{d^2V(x)}{dx^2} \right|_{{x=r_{p_0p_0'}}} \Delta
  r^2, \label{Vr}
\end{align}
where the expression for $\Delta r$ is given in equation~(\ref{Dr}). We finally obtain
\begin{align}
  \Delta E = \frac{\lambda^2}{r_0} \oint d \ell_0 \oint d \ell_0' \Big[ (h(\theta) +
  h(\theta')) \nonumber   \Big( V(r_{p_0p_0'})  \\  + \frac{r_{p_0p_0'}}{2}
  \frac{dV(r_{p_0p_0'})}{dr_{p_0p_0'}} \Big) + O(h^2) \Big]. \label{dE_1}
\end{align}
Now, we discuss the general case of $V(r) = \beta/r^{\gamma}$ ($\gamma$ is a
positive integer). While we focus on the Coulomb potential ($\gamma=1$) in this
work, scrutiny of the general case of $V(r)$ reveals the speciality of the
long-ranged Coulomb potential as will be shown. Up to the linear
term, we have 
\begin{align}
  \Delta E = \frac{(2-\gamma) \beta \lambda^2}{2r_0} \oint d \ell_0 \oint d \ell_0' \frac{h(\theta) +
  h(\theta')}{r_{p_0p_0'}^{\gamma}}.\label{DE}
\end{align}

Now, consider the perturbation in the form of 
\begin{align}
  r(\theta) = r_0 - \delta r_0 + h_0\cos n\theta, \label{n_mode}
\end{align} 
where the deformation mode is characterized by the nonzero
integer $n$. The shapes of the $n=1$ and $n=2$ modes are presented in
figure~\ref{deformation_final}(b); the red dashed reference circles are also shown. The
$\delta r_0$ term in equation~(\ref{n_mode}) (the zero mode) is introduced for the
conservation of the contour length of the ring, which leads to the relation
between $\delta r_0$, $h_0$ and $n$: $n^2 h_0^2=4 \delta r_0 (r_0-\delta r_0)$. 
By inserting equation~(\ref{n_mode}) into equation~(\ref{DE}), we have 
\begin{align}
  \frac{\Delta E}{2 - \gamma} &\propto& \int_{\Omega} d\omega   
  \frac{-  \delta r_0 + \frac{1}{4}h_0\cos(\frac{1}{2}nx)  \cos(\frac{1}{2}ny)   }{
    |\sin\frac{x}{2}|^{\gamma}},\nonumber \\  \label{DE_n}
\end{align}
where 
\begin{align}
\int_{\Omega} d\omega \equiv \int_{0}^{2\pi} dy \int_{-y}^{y} dx + \int_{2\pi}^{4\pi}
  dy \int_{-(4\pi-y)}^{4\pi-y} dx. \nonumber
\end{align}
The second term associated with $h_0$ in the integrand in equation~(\ref{DE_n}) is zero for nonzero
integers $n$ and $\gamma$. This conclusion is also true by replacing $h_0\cos
n\theta$ for $h_0\sum_m \cos m\theta$ in equation~(\ref{n_mode}), where $m$ is a
nonzero integer. It is important to notice that the sign of the remaining term
in equation~(\ref{DE_n}) depends on the value of $\gamma$. The marginal value for
$\gamma$ is 2 for the ring system, which is in contrast to the value of 4 for
the charged sphere system~\cite{jadhao2015coulomb}. This observation suggests
the connection of the instability of long-range interacting systems and the
dimension of space. For $\gamma > 2$, $\Delta E > 0$, indicating the stability
of the circular ring under relatively short-ranged interactions. However,
$\Delta E < 0$ at $\gamma=1$. This result indicates that the ground state shape
of the charged ring under Coulomb potential is not a perfect circle, which is
consistent with numerical observation.

\begin{figure}[t]  
\centering 
\includegraphics[width=3.3in]{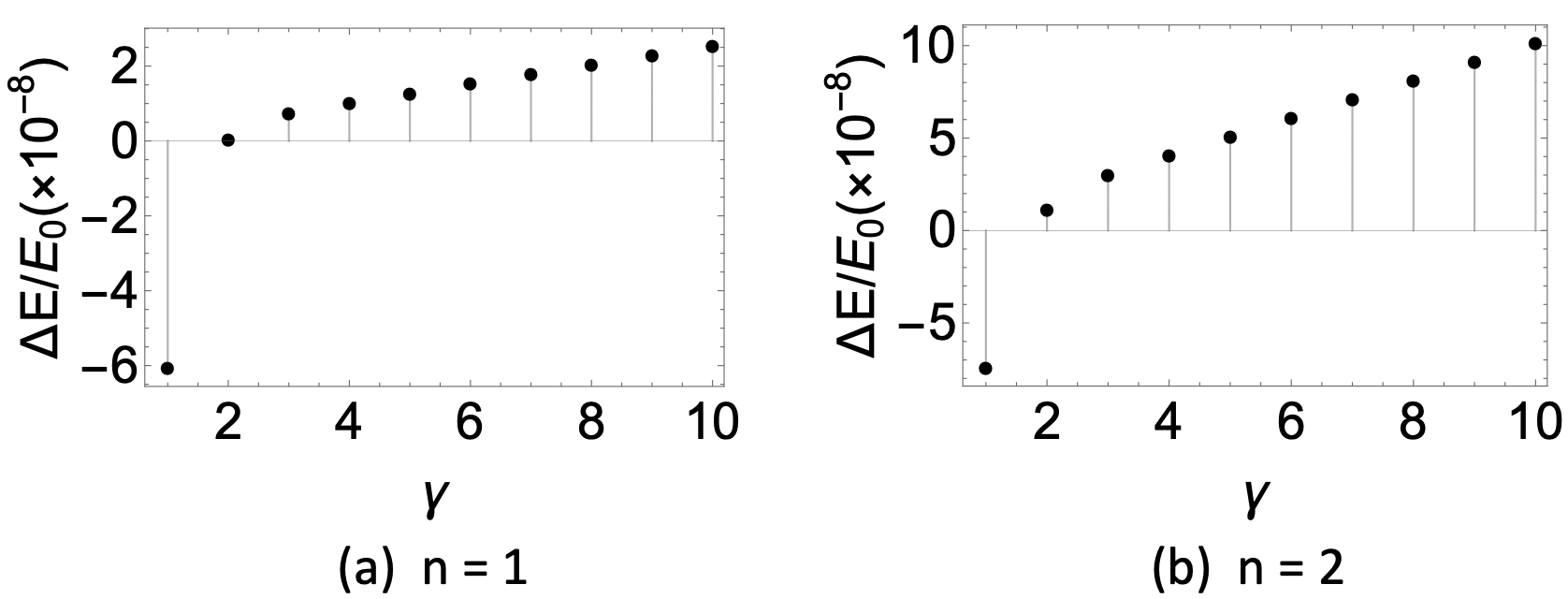}
  \caption{Plot of the change of the electrostatic energy $\Delta E$ by deforming
  a circular ring to the shape of the $n$-mode; see
  equation~(\ref{n_mode}) and figure~\ref{deformation_final}(b). $E_0$ is the total
  electrostatic energy of the unperturbed circular ring. The interaction potential takes the form of
  $V(r)\sim r^{-\gamma}$. $\delta r_0 = 10^{-2}$. $h_0=10^{-4}$. $r_0=1$.  $N=500$.
  These numerical data confirm the result of the analytical perturbation analysis.  }
\label{perturbation}
\end{figure}

We further perform numerical simulations to check the above analytical perturbation results.
Specifically, we place a collection of point charges evenly along the plane
curve of the $n$-mode shape specified by equation~(\ref{n_mode}), and compute the total electrostatic energy.
By comparison with the case of the corresponding circular ring of the same
contour length and charge density, we obtain the
change of the total electrostatic energy $\Delta E$ in the deformation. In
figure~\ref{perturbation}, we present the plots of $\Delta E$ versus $\gamma$ for the
$n=1$ and $n=2$ modes. Figure~\ref{perturbation} shows that $\Delta E<0$ for
$\gamma=1$, and $\Delta E>0$ for $\gamma>1$. These numerical results are
consistent with the analytical perturbation analysis. The small deviation of
$\Delta E$ from zero for $\gamma=2$ and $n=2$ in figure~\ref{perturbation}(b) shall be attributed to
the discrepancy of the continuous and the discrete charge distributions adopted
in the perturbation calculation and numerical simulations, respectively.
Examination of higher order terms ($n=3, 4, 5, 6$) also shows that $\Delta
E<0$ for $\gamma=1$, and $\Delta E>0$ for $\gamma>1$.

To conclude, the analytical perturbation calculation reveals the instability of
the circular shape of uniformly charged rings in the continuum limit. The
combination of the analytical perturbation analysis and the numerical results
shows that the ground state shape of the charged elastic ring is indeed not a
perfect circle.   While the $n=2$ mode has been identified as the lowest-energy
shape by the high-precision steepest descent method, we shall note that it is
still an open question to rigorously prove if this elliptic shape is the ground
state of the system. The subtlety of the long-range nature of the interaction
imposes challenges and brings richness to this question.

\begin{figure*}[t]  
\centering 
\includegraphics[width=6.75in]{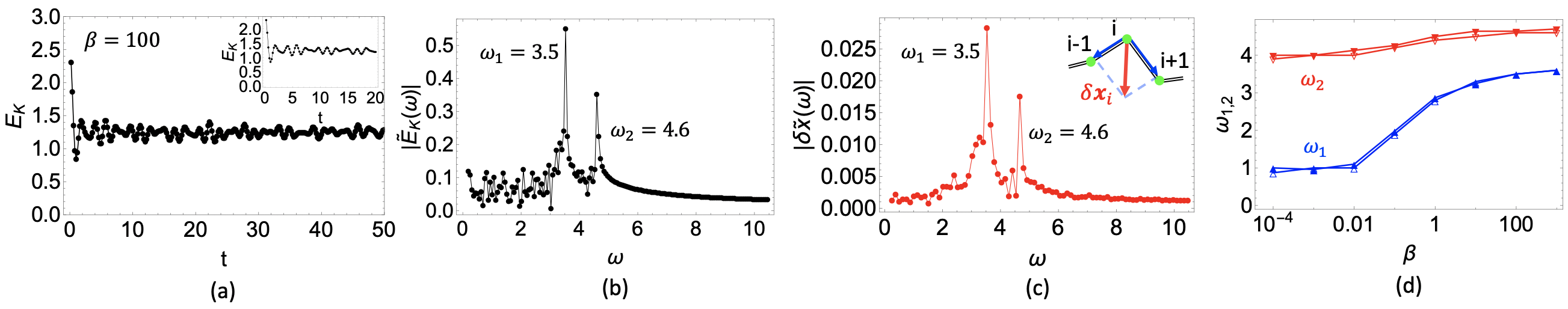}
  \caption{Spectral analysis of the kinetic energy of the randomly perturbed
  ring system. (a) Plot of the temporally-varying kinetic energy. $N=500$.
  $\beta=100$.  (b) The Fourier transformed kinetic energy curve in (a). (c) The
  Fourier transformed $\langle \delta x(t) \rangle$ curve. $\langle \delta x
  \rangle$ is the mean value of $\delta x_i$, which characterizes the relative
  motion of any particle $i$ with respect to its neighbors as shown in the
  inset. (d) Variation of the pair of the dominant frequencies with the strength
  $\beta$ of the electrostatic force.  $N=500$ (filled marks) and $N=1000$
  (empty marks). $v_0=0.1$. } 
  \label{dynamics}
\end{figure*}

\subsection{Dynamics of the ring}

The crucial role of the long-range electrostatic force in the static deformation
of the charged elastic ring system has been discussed in the preceding
subsections.  In this subsection, we proceed to discuss the dynamical effect of
the electrostatic force. To this end, we impose random disturbance by specifying
a random velocity field along the ring in the lowest-energy state, and analyze
the temporally-varying kinetic energy curve. The dynamical evolution of the ring
upon the disturbance is governed by the Hamiltonian dynamics. It is found that
the dynamical effect of the electrostatic force is reflected in the 
modulation of the frequency.  The relevant results are summarized in
figure~\ref{dynamics}.

Figure~\ref{dynamics}(a) shows the temporally-varying kinetic energy $E_K(t)$. The
total energy of the system is well conserved in numerical simulations; the
fluctuation of the total energy is at the order of $10^{-10}\%$. We perform
fourier transformation of the irregular kinetic energy curve, and
plot the $\tilde{E}_K(\omega)$ curve in figure~\ref{dynamics}(b).
Figure~\ref{dynamics}(b) shows that the dynamical rhythm of the interacting
point charges connected by linear springs is dominated by a pair of frequencies
$\omega_1$ and $\omega_2$. For comparison, we present the value
for the eigenfrequency $\omega_0$ of the elementary freestanding system consisting of two
identical masses $m$ connected by a linear spring of stiffness $k_0$. $\omega_0 =
\sqrt{2k_0/m}$, and the corresponding frequency of the kinetic energy curve is
doubled: $\omega_{E_k}=2\sqrt{2}\approx 2.8$.

To explore the physical origin of the dominant frequencies, we examine several
global and local dynamical modes, compute the relevant frequencies, and finally identify the mode
possessing the identical spectral structure as in the $\tilde{E}_K(\omega)$
curve. This mode corresponds to the relative motion of nearest neighboring
particles: $\delta \vec{x}_i= (\vec{x}_{i-1}-\vec{x}_{i}) +
(\vec{x}_{i+1}-\vec{x}_{i})$ [see the inset in figure~\ref{dynamics}(c)]. By
averaging over all of the particles and applying fourier transformation, we plot
the $\delta \tilde{x}(\omega)$ curve in figure~\ref{dynamics}(c). It turns out
that the frequencies at the two peaks in the $\delta \tilde{x}(\omega)$ curve in
figure~\ref{dynamics}(c) are identical to those in the $\tilde{E}_K(\omega)$
curve in figure~\ref{dynamics}(b). In general, oscillation of particles leads to
variation of energy. Here, we identify the long-range electrostatic force driven
short-wavelength dynamical mode (as characterized by $\delta \tilde{x}(\omega)$)
that could explain the origin of the pair of dominant frequencies in the
energy curve in figure~\ref{dynamics}(b).

We further find that the dominant frequencies, especially the value of
$\omega_1$, could be modulated by the strength $\beta$ of the
electrostatic force. Figure~\ref{dynamics}(d) shows the variation of the pair of
the dominant frequencies with the increase of $\beta$ for the cases of $N=500$
(filled marks) and $N=1000$ (empty marks). The insensitivity of the frequency
curves on the total number of particles could be attributed to the
short-wavelength nature of the dominant oscillation as presented in
figure~\ref{dynamics}(c). Here, we shall also emphasize the existence of the pair
of the dominant frequencies in a wide range of $\beta$ covering several orders of
magnitude. 

\

We finally briefly discuss the model of the charged elastic ring. The simplicity
of this model allows us to perform analytical perturbation calculation (under
certain limiting conditions). This model exhibits the 
deformation under the purely repulsive Coulomb potential among all of the
cases of $\gamma \in Z^+$ in the interaction potential $V(r)\sim
1/r^\gamma$, demonstrating the subtlety of the long-range force in the
organization of matter. Considering the significance of electrostatic force in
regulating materials at the nanoscale, the sinusoidal deformation phenomenon may
occur in small-scale structures in electrolyte solutions~\cite{Walker2011}. The
smallness of the intrinsic sinusoidal deformation imposes a challenge for direct
observation in real systems that are generally subject to thermal fluctuations.
In this aspect, the idealized charged elastic ring model shows its value in
revealing the subtle effect of long-range interaction.  This effect, as
demonstrated in the simple ring system, represents a
kind of complexity that is distinct from what is caused by many degrees of
freedom.

The ring model also serves as an example of shape instability under stretching.
In contrast to the well studied compression driven instabilities of various
elastic systems, the stretching caused instability under long-range repulsion
has received much less
attention~\cite{timoshenko1951theory,Landau1986,audoly2010elasticity}. Note that
the electromechanical stiffening effect has been reported in both elastic
membrane~\cite{ANDELMAN1995603} and rod~\cite{PhysRevE.67.061805,Rudnick2004}
systems.

Here, we may mention that the exploration of the instability phenomenon of the
simple ring model (the $S^1$ geometry) is inspired by our previously discovered
instability of the charged sphere system (the $S^2$
geometry)~\cite{jadhao2015coulomb}. To the best of our knowledge, the main
results of the charged elastic ring system (despite of its simplicity) presented
in this work have not been reported in previous work. Literature search shows
that, since the seminal work by Maxwell on the equilibrium distribution of
electric charges on a long narrow cylinder~\cite{maxwell1877electrical}, the
one-dimensional electrostatics problem has been extended from straight
lines~\cite{griffiths1996charge,jackson2000charge,jackson2002charge,amore2019thomson}
to circular~\cite{Ball1977,Zypman2006,Hafeez2009,Villa2019} and
eccentric~\cite{Denisova1998,amore2019thomson} rings of frozen geometry in both
continuum and discrete regimes. The electrostatic force on a closed, uniformly
charged space curve has been derived based on variational
principle~\cite{goldstein1995nonlinear,santiago2013elastic}. Statics and
dynamics of flexible ring systems have been studied for the uncharged
case~\cite{PhysRevE.64.011909}.

\section{Conclusion}

In summary, we investigated the deformation and dynamics of the classical
charged elastic ring system under the competing electrostatic and elastic
forces. Specifically, by the combination of high-precision numerical simulations
and analytical perturbation calculation, we showed the 
instability of the ring system, and the persistence of sinusoidal deformations
in the lowest-energy configurations. We also revealed the electrostatics-driven
elevation of the dominant frequencies in the dynamical evolution of the
perturbed rings. The study of the classical ring model advances our
understanding on the long-range nature of the physical interaction by its
manifestation in the organization and dynamics of matter.

\section*{Acknowledgements}

\noindent This work was supported by the National Natural Science Foundation of
China (Grants No. BC4190050).\\

\section*{References}

\providecommand{\newblock}{}

\bibliographystyle{iopart-num} 

\end{document}